\documentclass[twocolumn,showpacs,preprintnumbers,aps,amsmath,amssymb]{revtex4}

\usepackage[dvips]{graphicx}

\begin{document}

\title{Proposed Experimental Test for Proving Quantum Contextuality \\ with Non-entangled Photons}
\author{Yoshihiro Nambu}
\affiliation{Nano Electronics Research Laboratories, NEC, 34 Miyukigaoka, Tsukuba, Ibaraki 305-8501, Japan}

\begin{abstract}
We present a new and feasible test proving quantum contextuality in four-dimensional Hiltbert space. In our scheme, a contradiction between quantum mechanics and noncontextual hidden variables is revealed through the measurement statistics of two joint measurements of an ensemble of non-entangled qubits. The origin of nonclassicality is fully attributed to the measurement process. Quantum optical realization using photon pairs and linear optics are presented.
\end{abstract}


\maketitle

The foundation of quantum information technology (QIT) relies largely on the existence of non-classical phenomena that are classically unattainable. For example, entangled-states enable us to distribute secure random keys between two remote parties \cite{1}. Collective measurements enable us to break the Shannon limit for communication capacity using the signal coded in non-entangled states \cite{2}. The existence of non-classical phenomena is supported by the no hidden variable (no-HV) theorem \cite{3} of quantum mechanics (QM), which implies that in principle, there is no way of classically simulating the prediction of QM. The no-HV theorem refutes the HV model in combination with reasonable assumptions such as locality and noncontextuality \cite{3,4,4a}. These assumptions pose strict limitations on the HV model of QM, which implies the following no-HV theorems. Bell's theorem \cite{5} refutes the local HV (LHV) model of QM, which has been tested experimentally by many researchers \cite{6,7}. The Bell-Kochen-Specker (BKS) theorem \cite{5,8}, on the other hand, refutes the noncontextual HV (NCHV) model of QM. Non-contextuality (NC) implies that the value for an observable predicted by such a model does not depend on the experimental context, i.e., in which other co-measurable observables are measured simultaneously. 

Any experiment in the quantum context is made up of three processes; state-preparation, operation, and measurement. If we suppose just an experiment for which the HV model is consistent with QM, all the involved processes should be classical. In this way, one can define the classical processes. Bell's theorem is concerned with local measurements of an entangled state for revealing the nonlocal feature of QM. In this case, we attribute the origin of nonclasicality to state-preparation. On the other hand, the BKS theorem is related to the algebraic structure of quantum operators and is state-independent. In the original argument \cite{8}, the BKS theorem asserts that it is impossible to assign values to the squares of the spin components of a spin-1 particle along all directions on a sphere, subject to a sum rule for triads of orthogonal directions given by the total spin. At first glance, there seems to be no element of nonclassicality in their argument. However, their argument involves a measurement which reveals only the values of the squares of the spin components without revealing the values of the spin components themselves. This is a common nonclassical feature shared with the collective measurement \cite{4}. In recent years, experimental tests of quantum contextuality (QC) have been proposed by Cabello et al. \cite{9} and Simon et al. \cite{10} for a system with four-dimensional Hilbert space. Later, it was demonstrated by using two two-level degrees of freedom of single photons \cite{11} or single neutrons \cite{12,12a}. In these experiments, two co-measurable observables were analyzed with a sequence of compatible measurements (successive filtration) \cite{13} for an ensemble of two-qubit system initially prepared in a maximally entangled state. Both the state-preparation and measurement were nonclassical in their experiments. To the author's knowledge, no experimental test has been found that can prove the QC originated fully from the nonclassicality of the measurement according to the BKSfs argument.

In this letter, we show that a nonclassical state is unnecessary for experimental testing of QC by proposing an experiment using a classical state. Our proposal corresponds to the test of a certain state-specific version of the BKS theorem for a two-qubit system given by Peres \cite{14,14a} and Mermin \cite{13,15}. A contradiction between QM and NCHV can be demonstrated in the measurement statistics of two different joint measurements by using an ensemble of a two-qubit system initially prepared in a specific product state. We also present a quantum optical implementation. 

Before describing our proposal, we briefly recall the argument given by Mermin \cite{15} and clarify the contradiction between QM and NCHV for a two-qubit system. Any observable of a single qubit can be represented by the linear combination of the set of operators $\{\sigma_{0}^{i},\sigma _{1}^{i},\sigma _{2}^{i},\sigma _{3}^{i}\}$, where  $\sigma _{0}^{i}$  and  $\sigma _{\mu}^{i}$  (${\mu} =1, 2, 3$) are the identity and Pauli operators for the subsystem \textit{i} (\textit{i} = 1, 2), respectively. Then, one can easily confirm the following commutation relations for the product Pauli-operators: 
\begin{equation}
[\sigma_{2}^{1} \otimes \sigma_{2}^{2}, \sigma_{3}^{1} \otimes \sigma_{3}^{2}]=[\sigma_{2}^{1} \otimes \sigma_{3}^{2}, \sigma_{3}^{1} \otimes \sigma_{2}^{2}]=0
\label{eq1}
\end{equation}
and the following operator identity \cite{4,15}
\begin{equation}
-(\sigma_{2}^{1} \otimes \sigma_{2}^{2})(\sigma_{3}^{1} \otimes \sigma_{3}^{2})
= (\sigma_{2}^{1} \otimes \sigma_{3}^{2})(\sigma_{3}^{1} \otimes \sigma_{2}^{2})
= \sigma_{1}^{1} \otimes \sigma_{1}^{2}.
\label{eq2}
\end{equation}
These results lead to contradicting predictions between QM and NCHV for a feasible empirical test. To observe this, we compare two predictions. The major assumptions in an NCHV theory are as follows. First, every observable of an individual system possesses a definite preexisting value waiting to be revealed by any appropriate measurement, which must be one of its eigenvalues. Second, the values associated with every subset of mutually commuting observables obey certain algebraic identities obeyed by the observables themselves (functional consistency) \cite{13,15}. It implies that 
\begin{eqnarray}
v[(\sigma_{2}^{1} \otimes \sigma_{2}^{2})(\sigma_{3}^{1} \otimes \sigma_{3}^{2})] 
 &=& v(\sigma_{2}^{1} \otimes \sigma_{2}^{2}) v(\sigma_{3}^{1} \otimes \sigma_{3}^{2}) \nonumber \\
 &=& v(\sigma_{2}^{1})v(\sigma_{2}^{2})v(\sigma_{3}^{1})v(\sigma_{3}^{2}),
\label{eq3}
\end{eqnarray}
where $v(A)$ is a numerical value for observable $A$ possessed by an individual system. A similar equation holds for $v[(\sigma_{2}^{1} \otimes \sigma_{3}^{2})(\sigma_{3}^{1} \otimes \sigma_{2}^{2})]$. Eventually, we obtain 
\begin{equation}
v[(\sigma_{2}^{1} \otimes \sigma_{3}^{2})(\sigma_{3}^{1} \otimes \sigma_{2}^{2})]=v[(\sigma_{2}^{1} \otimes \sigma_{2}^{2})(\sigma_{3}^{1} \otimes \sigma_{3}^{2})].
\label{eq4}
\end{equation}
Note that Eqs. (\ref{eq2}) and (\ref{eq4}) already imply a contradiction between QM and NCHV in assigning values for the commutating observables to a single individual system. However, this contradiction can not be empirically testable for a individual system since these equations involve noncompatible observables that cannot be measured in the same individual system.

The counterfactual difficulty involved in the above argument has been usually resolved in the following way. We prepare an ensemble of an identical system and evaluate its noncompatible observables by changing the experimental configuration. We discuss an empirically testable quantity that is given by the values of the observables averaged over a given ensemble under this situation. Let us consider the quantity 
\begin{equation}
C=\langle (\sigma_{2}^{1} \otimes \sigma_{3}^{2})(\sigma_{3}^{1} \otimes \sigma_{2}^{2}) \rangle - \langle(\sigma_{2}^{1} \otimes \sigma_{2}^{2})(\sigma_{3}^{1} \otimes \sigma_{3}^{2})\rangle, 
\label{eq5}
\end{equation}
where $\langle \cdots \rangle $ indicates the average over an ensemble of the two-qubit systems in the given state. From Eq. (\ref{eq2}), standard QM predicts that the value of $C$ should be $C^{\textrm{QM}}=2 \langle \sigma_{1}^{1} \otimes \sigma_{1}^{2} \rangle $. On the contrary, NCHV implies that $C^{\textrm{NCHV}}=0$, since Eq. (\ref{eq4}) implies that two quantities in the parentheses of Eq. (\ref{eq5}) should have the same statistics for a given state. Therefore, if we evaluate $C$ for the system in a state with $\langle \sigma_{1}^{1} \otimes \sigma_{1}^{2} \rangle \neq 0$, we can experimentally demonstrate the contradiction between QM and NCHV from the measurement statistics. Previous research \cite{10,14} proved an inconsistency in assigning the noncontextual values to the nine observables $\sigma_{2}^{1}$, $\sigma_{2}^{2}$, $\sigma_{3}^{1}$, $\sigma_{3}^{2}$, $\sigma_{2}^{1} \otimes \sigma_{2}^{2}$, $\sigma_{3}^{1} \otimes \sigma_{3}^{2}$, $\sigma_{2}^{1} \otimes \sigma_{3}^{2}$, $\sigma_{3}^{1} \otimes \sigma_{2}^{2}$, and $\sigma_{1}^{1} \otimes \sigma_{1}^{2}$ for the simultaneous eigenstate of $\sigma_{2}^{1} \otimes \sigma_{2}^{2}$ and $\sigma_{3}^{1} \otimes \sigma_{3}^{2}$. In contrast, the key point of the present proposal is to prove a similar inconsistency for the eigenstate of $\sigma_{1}^{1} \otimes \sigma_{1}^{2}$. Each of these proofs constitutes a multiplicative proof of the state-specific BKS theorem. They can be converted into a standard proof of the state-specific BKS theorem associated with different given states \cite{14a}. The former includes 15 rays in $\boldsymbol{R}^4$ whereas the latter 20 rays in $\boldsymbol{R}^4$.

To evaluate the value of $C$ experimentally, we need to perform two kinds of joint measurements; one is the joint measurement of $\sigma_{2}^{1} \otimes \sigma_{2}^{2}$ and $\sigma_{3}^{1} \otimes \sigma_{3}^{2}$, and the other is that of $\sigma_{2}^{1} \otimes \sigma_{3}^{2}$ and $\sigma_{3}^{1} \otimes \sigma_{2}^{2}$. We consider a sequential measurement strategy proposed by Simon \cite{10}. In the following, the joint measurement of the former pair of observables is described in detail as an illustrative example. 

\begin{figure}[b]
\begin{center}
\includegraphics[scale=0.8]{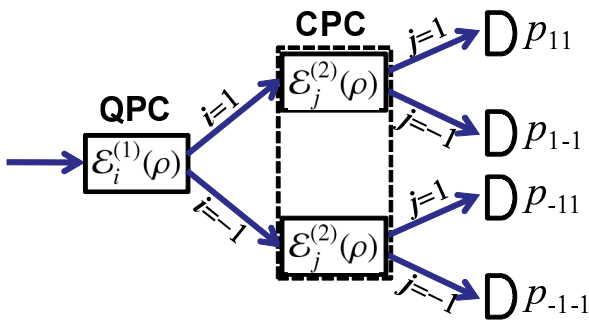}
\caption{Successive filtering strategy for performing joint measurements of $\sigma_{2}^{1} \otimes \sigma_{2}^{2}$ and $\sigma_{3}^{1} \otimes \sigma_{3}^{2}$. $p_{ij}$ stands for the probability of finding the two qubits after first filtration with parity $i$ and second filtration with parity $j$.}
\label{F1}
\end{center}
\end{figure}

Consider operations $\mathcal{E}_{i}^{(1)}$ and $\mathcal{E}_{j}^{(2)}$, which depend on the parity ($i, j=\pm 1$) of the input qubits and are applied sequentially on a two-qubit system (Fig. \ref{F1}). The first operation describes the filtration associated with the measurement of $\sigma_{3}^{1} \otimes \sigma_{3}^{2}$ and written as 
\begin{equation}
\mathcal{E}_{\pm 1}^{(1)}(\rho)=(1-\frac{p}{2}) P_{\pm 1} \rho P_{\pm 1}+\frac{p}{2}Q_{\pm 1}\rho Q_{\pm 1},
\label{eq6}
\end{equation}
where ${\rho}$ is the state of the input two-qubit system, $p (0 \leq p \leq 1)$ is the mixing parameter, $P_{\pm 1}$ is the projector onto a two-dimensional subspace spanned by the eigenstates of $\sigma_{3}^{1} \otimes \sigma_{3}^{2}$ with eigenvalue $\pm 1$; 
\begin{equation}
P_{\pm 1}=\frac{1}{2}(\sigma_{0}^{1} \otimes \sigma_{0}^{2} \pm \sigma_{3}^{1} \otimes \sigma_{3}^{2}),
\label{eq7}
\end{equation}
and $Q_{\pm 1}$ is the operator introduced to incorporate a specific type of non-ideality into the considered operation; 
\begin{equation}
Q_{\pm 1}=\frac{1}{2}(\sigma_{0}^{1} \otimes \sigma_{3}^{2} \pm \sigma_{3}^{1} \otimes \sigma_{0}^{2}).
\label{eq8}
\end{equation}
This describes the general trace-decreasing operation for filtering the two-qubit systems with a predefined parity $i$ both defined in the $\sigma_{3}$ basis. For example, the operation with a plus sign transfers the qubits only if the first and second qubits have the same values in $\sigma_{3}$ (even parity), whereas one with a minus sign transfers only if the two qubits have different values (odd parity). This general operation encompasses two specific filtering operations for special cases. When $p=0$, the operation is referred to as the quantum parity check (QPC) \cite{17,18,19}. This operation is non-classical in which the parity of the qubits is verified without measuring or determining the value of either qubit. On the other hand, when $p=1$, the operation is entirely classical and referred to as the classical parity check (CPC), which describes the filtering strategy of the two-qubit system with a predefined parity by local filtration of either qubit (like the Stern-Gelrach apparatus) associated with the predefined value of the parity. This task requires classical communication concerning the pointer value of the local filtration associated with every filtration event.

On the other hand, the second filtration is associated with the measurement of $\sigma_{2}^{1} \otimes \sigma_{2}^{2}$ and written as 
\begin{subequations}
\begin{eqnarray}
\mathcal{E}_{+1}^{(2)}(\rho)&=&P_{LL} \rho P_{LL}+P_{RR} \rho P_{RR}, \\
\label{eq9a}
\mathcal{E}_{-1}^{(2)}(\rho)&=&P_{LR} \rho P_{LR}+P_{RL} \rho P_{RL}
\label{eq9b}
\end{eqnarray}
\end{subequations}
where $P_{xy} (x,y=L,R)$ is the products of the projectors onto a one-dimensional subspace spanned by the eigenstates of $\sigma_{2}$; 
\begin{subequations}
\begin{equation}
P_{xy}=P_{x}^{1} \otimes P_{y}^{2},
\label{eq10a}
\end{equation}
with
\begin{eqnarray}
P_{L}^{i}&=&\frac{1}{2}(\sigma_{0}^{i}+\sigma_{2}^{i}),\\
\label{eq10b}
P_{R}^{i}&=&\frac{1}{2}(\sigma_{0}^{i}-\sigma_{2}^{i}).
\label{eq10c}
\end{eqnarray}
\end{subequations}
This operation is nothing but a CPC for filtering the two-qubit systems with a predefined parity $i$ both defined in the $\sigma_{2}$ basis. 

The value of $C$ can be experimentally obtained in the following procedure. We first evaluate the probability $p_{ij}$ of finding two qubits initially prepared in a state $\rho$ after successive filtrations $\mathcal{E}_{i \wedge j}=\mathcal{E}_{j}^{(2)}  \circ \mathcal{E}_{i}^{(1)}$. The probability is given by 
\begin{equation}
p_{ij}=Tr_{12}\mathcal{E}_{i \wedge j}(\rho).
\label{eq11}
\end{equation}
It is easy to show that when $p=0$,
\begin{equation}
\langle (\sigma_{2}^{1} \otimes \sigma_{2}^{2})(\sigma_{3}^{1} \otimes \sigma_{3}^{2}) \rangle=p_{11}+p_{-1-1}-p_{-11}-p_{1-1}.
\label{eq12}
\end{equation}
Therefore, we can evaluate the second term of the right-hand portion of Eq. (\ref{eq5}) by evaluating the probabilities $p_{ij}$ of finding two qubits for all the pairs of $(i,j)$. Note that it is necessary to use an exact QPC in the first stage of the filtrations to obtain the exact value of the ensemble average of the joint observable considered. Second, we evaluate similar probabilities $p'_{ij}$ of finding two qubits for the same initial state $\rho$ after successive filtrations $\mathcal{E'}_{i \wedge j}=\mathcal{E}_{j}^{(4)}  \circ \mathcal{E}_{i}^{(3)}$ to obtain the value of the first term of the right-hand portion of Eq. (\ref{eq5}). Here, $\mathcal{E}_{i}^{(3)}$ is a modified QPC, which only filters two-qubit systems with a predefined parity defined in the $\sigma_{2}$ and $\sigma_{3}$ bases, and $\mathcal{E}_{j}^{(4)}$ is a modified CPC, which only filters two-qubit systems with a predefined parity defined in the $\sigma_{3}$ and $\sigma_{2}$. Finally, the difference of these values gives the value of $C$. 

To obtain the maximal contradiction between QM and NCHV, we only need to prepare the initial state with $|\langle \sigma_{1}^{1} \otimes\sigma_{1}^{2} \rangle |=1$ for this successive filtration measurement. Such a state can be easily prepared. For example, we can choose the product state that is the eigenstate of $\sigma_{1}^{1} \otimes\sigma_{1}^{2}$
\begin{eqnarray}
\lefteqn{\rho_{0} = \left| ++ \right\rangle \left\langle ++ \right| } \nonumber \\
 &=& \frac{1}{4}(\sigma_{0}^{1} \otimes\sigma_{0}^{2} + \sigma_{0}^{1} \otimes \sigma_{1}^{2} + \sigma_{1}^{1} \otimes \sigma_{0}^{2} + \sigma_{1}^{1} \otimes \sigma_{1}^{2}),
\label{eq13}
\end{eqnarray}
where $|+ \rangle =\sigma_{1}|+ \rangle$ is the eigenstate of $\sigma_{1}$ with eigenvalue one. In fact, it is easy to see that $C=2-p-p'$ holds, where we assumed different mixing parameters $p$ and $p'$ for $\mathcal{E}_{i}^{(1)}$ and $\mathcal{E}_{i}^{(3)}$, and $C$ is estimated using Eq.(\ref{eq12}) despite the fact that it is applicable only when $p=p'=0$. As expected, $C$ approaches 2 if both $p$ and $p'$ approach zero. Therefore, to disprove NCHV experimentally, we need not use an entangled state.

In spite of the above statement, we actually have an entangled state during the measurement process unless $p=1$, which is apparently non-classical. For example, the post-measurement state just after first stage of the filtrations of $\mathcal{E}_{1}^{(3)}$ is

\begin{equation}
\mathcal{E}_{1}^{(3)}(\rho_{0})=\frac{1}{2}\{(1-\frac{p}{2})|{\tilde {\Phi}}^{+}\rangle \langle {\tilde {\Phi}}^{+}|+ \frac{p}{2} |{\tilde {\Phi}}^{-}\rangle \langle {\tilde {\Phi}}^{-}|\},
\label{eq14}
\end{equation}
where $| \Phi^{+} \rangle$, $| \Phi^{-} \rangle$, $| {\tilde {\Phi}}^{+} \rangle=U_{1}(\frac{\pi}{2}) \otimes \sigma_{0}^{2}| \Phi^{+} \rangle$ and $| {\tilde {\Phi}}^{-} \rangle =U_{1}(\frac{\pi}{2}) \otimes \sigma_{0}^{2}| \Phi^{-} \rangle$ are the two Bell states and their unilateral rotation around the 1-axis. These post-measurement states depend on the filtering operation for measuring the observable unless $p=1$ which is the sources of QC. Thus, our test of QC is considered to be a manifestation of the non-classicality of QM in the measurement process. In the test of Bell's theorem, non-classicality in the non-local state was revealed using the local measurements. In the present test of QC, the prepared state is definitely classical, but a collective measurement was used to reveal the non-classicality.

\begin{figure}[t]
\begin{center}
\includegraphics[scale=0.6]{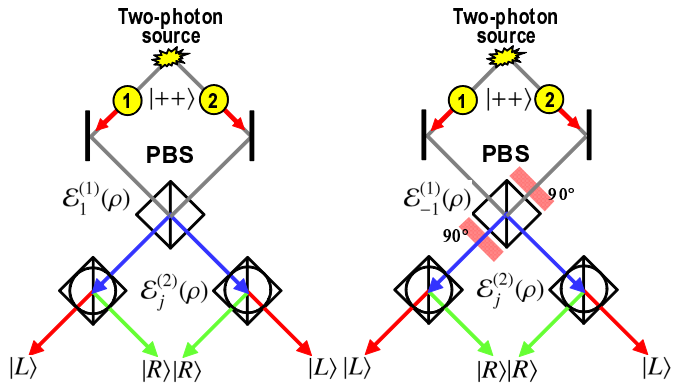}
\caption{Experimental setup for proving quantum contextuality. $|+\rangle =U_{2}(-\frac{\pi}{2})|0 \rangle $, $|L \rangle=U_{1}(\frac{\pi}{2})|0 \rangle $ and so on, and $|0 \rangle$ stands for horizontally polarized single photon state.}
\label{F2}
\end{center}
\end{figure}

Let us now turn our attention to the implementation of the present test of QC by using qubits encoded in the polarization state of a single photon. We adopt the isomorphism $| 0 \rangle  \equiv | H \rangle$, $| 1 \rangle  \equiv | V \rangle$, where $| H \rangle$, $| V \rangle$ denote the horizontal and vertical polarizations, respectively. Figure \ref{F2} illustrates an experimental setup using photon pairs produced by, say, parametric down conversion and linear optics. The first stages of the filtration, $\mathcal{E}_{\pm 1}^{(1)}$ and $\mathcal{E}_{\pm 1}^{(3)}$, can be effectively implemented by using a polarizing beam splitter (PBS) and post selection based on the coincidence detection of photons behind its two outputs \cite{17,18,19}. The PBS transmits the horizontal polarization and reflects the vertical one. If a single photon is injected into each input of the PBS and a single photon is emitted from each of the two outputs, it implies that the injected photons have the same polarization, i.e., the same parity, and are both simultaneously transmitted through or reflected at the PBS. Otherwise, either both or neither of the photons emit from the two outputs of the PBS. Successful implementation of the parity check operation corresponds to the registration of a photon for two detectors each in two downstream channels of the PBS; otherwise the operation fails. In addition, if two photons are indistinguishable and the propagated path of the two photons cannot be addressed in principle, $p=0$ is attained. In this case, successful operation projects the input state onto a two-dimensional subspace spanned by $\left|00\right\rangle$ and $\left|11\right\rangle$, which is isomorphic to the projector $P_{+1}$ (left-hand figure). Note that for this to work, we need to eliminate the imbalance in the two paths carefully leading to which-path information. The operation represented by the complementary projector $P_{-1}$ can be implemented exclusively by adding $90^\circ$ rotation to one pair of the input and output modes by using half-wave plates as shown in the setup in the right-hand figure. Therefore, these setups would be satisfactory in the operation associated with $\mathcal{E}_{\pm 1}^{(1)}$. Satisfactory operation associated with $\mathcal{E}_{\pm 1}^{(3)}$ will also occur if we add quarter-wave plates at the appropriate position in the path. On the other hand, the second stages of the filtration, $\mathcal{E}_{\pm 1}^{(2)}$ and $\mathcal{E}_{\pm 1}^{(4)}$, can be easily implemented using quarter-wave plates and PBSs, downstream from the first PBS, and photon coincidence detection circuits. By analyzing the coincidence counting statistics, we can easily obtain the probabilities given in Eq. (\ref{eq12}) to calculate $C$. It should be noted that it is crucial to assign the dichotomic variable \{-1, 1\} to the real polarization consistently throughout the experiment in order to obtain the correct value of $C$.

In conclusion, we proposed a new and feasible empirical test of QC using photon pairs and collective operation. Maximal contradiction between QM and NCHV can be observed using photon pairs in an appropriate product polarization state, linear optics, and photon coincidence measurements. We have illustrated, for the first time, that QC is empirically testable from the measurement statistics of two different joint measurements to which the origin of non-classicality is fully attributed as in the original argument of the BKS theorem. 

We would like to thank Yuta Okubo, Kunihiro Kojima, and Akihisa Tomita for their helpful discussions. We also thank Yutaka Shikano and Yuji Hasegawa for their comments. This work was supported by the National Institute of Information and Communication of Technology, Japan.

\end{document}